\documentclass[12pt]{article}
\usepackage{graphicx}
\usepackage{epsfig}
\usepackage{epstopdf}
\DeclareGraphicsExtensions{.pdf,.eps,.png,.jpg,.mps}

\def\lsim{\mathrel{\rlap{\lower3pt\hbox{\hskip0pt$\sim$}}
     \raise1pt\hbox{$<$}}}         
\def\gsim{\mathrel{\rlap{\lower4pt\hbox{\hskip1pt$\sim$}}
     \raise1pt\hbox{$>$}}}         

\usepackage{amsmath}
\usepackage{amsfonts}

\begin{document}
\begin{titlepage}

\centerline{\Large \bf Brane World}
\medskip

\centerline{Zura Kakushadze$^\S$$^\dag$$^\ddag$\footnote{\, Email: \tt zura@quantigic.com}}
\bigskip

\centerline{\em $^\S$ Quantigic$^\circledR$ Solutions LLC}
\centerline{\em 1127 High Ridge Road \#135, Stamford, CT 06905\,\,\footnote{\, DISCLAIMER: This address is used by the corresponding author for no
purpose other than to indicate his professional affiliation as is customary in
publications. In particular, the contents of this paper
are not intended as an investment, legal, tax or any other such advice,
and in no way represent views of Quantigic® Solutions LLC,
the website \underline{www.quantigic.com} or any of their other affiliates.
}}
\centerline{\em $^\dag$ Department of Physics, University of Connecticut}
\centerline{\em 1 University Place, Stamford, CT 06901}
\centerline{\em $^\ddag$ Free University of Tbilisi, Business School \& School of Physics}
\centerline{\em 240, David Agmashenebeli Alley, Tbilisi, 0159, Georgia}
\medskip
\centerline{(October 5, 2014)\footnote{\, This article (in English), with some differences ({\em e.g.}, the Higgs boson had not been discovered yet), appeared some years ago in an online magazine Kvali.com.}}

\bigskip
\medskip

\begin{abstract}
{}This article attempts to discuss some key aspects of the Brane World scenario in a popular fashion, without using any formulas.
\end{abstract}
\bigskip
\medskip

\noindent{}{\bf Keywords:} Brane World, string theory, Standard Model, gravity, grand unification

\end{titlepage}

\newpage

{}Humans have been curious about nature since prehistoric times. Why does rain fall? Why does fire burn? Why is the sky blue? The desire to answer such questions necessitated addressing other, more and more fundamental ones, which gradually crystallized into physics. Physics is a science that at its deepest levels deals with fundamental laws governing nature. Theoretical physics uses mathematics as a language for precisely formulating laws of nature and explaining or sometimes even predicting various observable phenomena. Experimental physics develops tools and technology for discovering new phenomena or verifying (and sometimes ruling out) various theoretical predictions.

{}Presently theoretical physics is substantially ahead of its experimental counterpart -- human imagination has gone far beyond our technological abilities to discover more fundamental underlying structures. In this sense most developments in theoretical physics that occurred in, say, the past few decades are somewhat speculative as they have not yet been experimentally verified or ruled out. Nonetheless, many of these novel theories are very intriguing and fascinating and even promising in terms of their experimental verification (or exclusion) in the foreseeable future. In this article we attempt to review some of these developments in a popular fashion, that is, without using any of the complex mathematical apparatus that physicists normally utilize in their discussions on the subject.

{}Let us start with the current paradigm, which is referred to as the Standard Model of elementary particles. Within this experimentally verified framework, everything we see (as well as many things we do not) around us is made of point-like particles, which are the fundamental building blocks for all matter. Thus, for instance, water consists of water molecules, and a water molecule is made of two atoms of hydrogen and one atom of oxygen. Atoms themselves are not the most fundamental building blocks either. An atom consists of electrons ``circling" around a nucleus. Within the Standard Model electron is a fundamental point-like particle and is a member of a larger group of such objects referred to as leptons. On the other hand, nuclei have further substructure; they are made of protons and neutrons, which are members of a larger group of such objects referred to as baryons, which in turn are made of fundamental point-like particles called quarks.

{}Quarks and leptons interact with each other via four known forces of nature. All matter is subject to the gravitational interaction whose discussion we postpone until a bit later. The other three forces go as follows. Electrons carry electric charge, so do some (but not all -- see below) leptons as well as all quarks. All electrically charged particles interact with each other via the electromagnetic interaction (thus, two particles with same sign charges repel, while those with opposite sign charges attract). This interaction occurs via electrically charged particles exchanging with each other the so-called photons, which are point-like particles responsible for transmitting the electromagnetic interaction. Photons are electrically neutral, in fact, they carry no other charges. They propagate in vacuum with the speed of light (this is because they are massless), so the electromagnetic interaction is a long-range interaction (thus, it can readily be detected at everyday distance scales, not just at subatomic ones). Actually, light as we know it is nothing but a collection of photons. Thus, photons are quanta (or the building blocks) of the electromagnetic interaction. In fact, all four interactions have such point-like quanta.

{}As we mentioned above, some leptons are electrically neutral. Such leptons are called neutrinos. They do not interact with electrons (or each other) via the electromagnetic interaction, but one type of neutrinos (the so-called left-handed neutrinos; there are also believed to exists the so-called right-handed neutrinos that interact with other matter only via the gravitational interaction) do interact with each other as well as some other leptons via the so-called weak interaction. The quanta of the weak interaction are referred to as W- and Z-bosons. They differ from photons in two important ways. First, the W- and Z-bosons are massive, so they transmit the weak interaction to short subatomic distance scales (this is why detecting the weak interaction is substantially more difficult compared with the electromagnetic interaction). Second, there are actually two W-bosons, which carry opposite electric charges. They are also oppositely charged under the so-called weak isospin, which plays a role for the weak interaction analogous to that of the electric charge for the electromagnetic interaction. On the other hand, there is only one Z-boson, which carries no electric or weak isospin charges. In this regard the Z-boson resembles the photon, but unlike the photon, which is massless, the Z-boson is massive.

{}All quarks are electrically charged (in fact, they have electric charges which are multiples of 1/3 of the electron charge) and are subject to the electromagnetic interaction. They can also interact with each other via the weak interaction. However, unlike leptons, quarks are subject to yet another interaction referred to as the strong interaction. Thus, quarks carry another type of charge called color (this is just a flashy name not to be confused with color we recognize in everyday life). There are three different colors, and the strong interaction between quarks is transmitted via exchanging the corresponding quanta called gluons. There are eight different gluons labeled by different combinations of color. In fact, color plays a role for the strong interaction analogous to that of the weak isospin for the weak interaction. In this regard gluons resemble W- and Z-bosons. Unlike the latter, and more like photons, however, gluons are massless. At first this might suggest that the strong interaction should be a long-range interaction just like the electromagnetic one. However, the strong interaction is stronger (hence the name) than the electromagnetic interaction by about two orders of magnitude, so quarks and gluons are trapped inside of nuclei under normal conditions (albeit under some special circumstances it is possible to have quark-gluon plasma where quarks and gluons no longer form nuclei). Let us mention that leptons do not carry color charges, nor do gluons carry electric or weak charges, so leptons and gluons talk to each other only via the gravitational interaction. On the other hand, quarks and leptons can interact with each other via the electromagnetic and weak interactions.

{}The fact that photons (quanta of the electromagnetic interaction) and gluons (quanta of the strong interaction) are massless, while W- and Z-bosons (quanta of the weak interaction) are massive might appear to be a bit curious. Thus, the electromagnetic and weak interactions are manifestations of the underlying unified electroweak interaction, and the latter has four massless quanta. However, the electroweak interaction is broken into the electromagnetic and weak interactions because W- and Z-bosons interact with the so-called Higgs boson, which is another fundamental particle, while the photon does not. This interaction results in W- and Z-bosons acquiring masses, while the photon remains massless. This is referred to as the spontaneous breaking of the electroweak symmetry. The full electroweak symmetry is actually restored at high energy scales where the W- and Z-boson masses can be neglected (that is, where they can be treated as being effectively massless).

{}The fact that the electromagnetic and weak interactions are unified into the electroweak interaction is suggestive of the Grand Unification idea where one attempts to unify the remaining three forces of nature, that is, the electroweak, strong as well as gravitational interactions. Let us note that this is beyond the scope of the Standard Model, which has been experimentally verified. In particular, energies required to test grand unified models are beyond the capabilities of existing colliders (huge experimental machines where properties of fundamental particles and their interactions are tested).  So the discussion below is based on yet to be verified (or ruled out) theoretical ideas, and is therefore largely speculative.

{}The idea of Grand Unification is very attractive and intriguing, and various grand unified models have been proposed. Albeit there are some non-trivial issues in such model building, let us discuss some key conceptual aspects of it. One can readily unify the electroweak and strong interactions. Recall that there are four electroweak quanta and eight strong quanta. In the simplest scenario one has a grand unified interaction with 24 massless quanta, 12 of which remain massless after the corresponding spontaneous symmetry breaking (these are the twelve quanta of the electroweak and strong interactions), while the other 12 (the so-called X- and Y-bosons) acquire masses via interactions with the corresponding Higgs bosons. The masses of these 12 heavy quanta are so large that they cannot be discovered in the existing collider experiments. Such grand unified models reduce the number of the fundamental interactions to two -- a grand unified interaction and the gravitational interaction. Can we unify them further into one most fundamental interaction?

{}This is where a big conceptual problem comes in. The gravitational interaction is actually very different from the electromagnetic, weak and strong interactions. It is (believed to be) transmitted by the corresponding quanta, which are called gravitons and propagate with the speed of light (that is, gravitons are massless). In this regard gravitons resemble quanta of other interactions. However, they are very different from the latter in one very important way, which brings us to the concept of spin.

{}Spin is a fundamental property of elementary particles. In particular, it is conserved in interactions. In particle physics conventions spin is always a multiple of 1/2. Thus, quarks and leptons carry 1/2 spin, while photons, W- and Z-bosons as well as gluons carry spin 1. Higgs bosons carry spin 0. Thus, spin does not obstruct unification of electromagnetic, weak and strong interactions into one interaction as all of the corresponding quanta carry spin 1. On the other hand, Higgs bosons cannot be unified with, say, photons in this fashion - they carry different spin (actually, one can unify particles with different spin via the so-called supersymmetry; however, supersymmetry alone does not suffice to unify interactions). Now, gravitons carry spin 2, not spin 1, which makes them very different from other interaction quanta, and poses a major obstacle in unifying the gravitational interaction with the other interactions.

{}All attempts to circumvent the aforementioned difficulty in unifying gravitational and other known interactions within the framework of point-like elementary particles had failed. However, another framework, which is called string theory, appears to allow such unification. In this framework fundamental objects are not point particles but strings.

{}In string theory point-like particles are not the fundamental building blocks. Rather, various elementary particles are manifestations of different vibration modes of the same fundamental string. Here we can draw an analogy with the everyday life strings. Thus, for instance, consider a guitar string. This string has the so-called fundamental tone, which is the longest wavelength (or the lowest frequency) tone, as well as an infinite number of shorter wavelength (or higher frequency) tones, the so-called overtones. Various elementary particles are then analogous to different tones of the guitar string. What is fascinating about string theory is that the exact same fundamental string has vibration modes with different spin. In fact, a consistent string theory always possesses a massless mode with spin 2, the graviton. That is, gravitational interaction is always part of string theory. There are believed to be five consistent string theories, and some of them contain spin 1 interaction quanta necessary to accommodate the other three known interactions. So, in principle, one could say that string theory could be the correct framework for (perhaps surprising) unification of all known fundamental interactions.

{}However, making a connection between string theory and the real world is not so easy. And there are various reasons why. First, all known consistent string theories live in 10 space-time dimensions (with 9 space-like and 1 time-like dimensions), while the real world is 4-dimensional (with 3 space-like and 1 time-like dimensions). More precisely, interactions in string theory are consistent only in 10 dimensions assuming that all of the dimensions where strings propagate are infinite. One way of reducing the number of these dimensions is to assume that the extra 6 space-like dimensions are curled up or compactified. That is, the extra dimensions are very small (for instance, small circles), and cannot be detected using current experimental tools. There are two issues with this approach. First, one of the robust predictions of string theory is that there are more than 4 dimensions. If these dimensions are so small that we cannot even observe them, it substantially reduces the usefulness of this prediction (or at least postpones its usefulness to possibly very distant future) -- imagine if we could discover precisely 6 extra dimensions, this would certainly smell like a ``smoking gun" for string theory. Second, the number of consistent ways for compactifying these extra dimensions appears to be so large (actually, it is infinite) that the number of 4-dimensional theories resulting from string theory is enormous, and most of these 4-dimensional theories do not even remotely look like the real world. Finding the one, if any, corresponding to the real world is then like looking for a needle in a haystack. String theory in this context does not appear to be any more predictive then the theory of point-like particles it replaces.

{}Another difficulty is related to the fact that, at least in its traditional formulation, to truly test string theory, in other words, to start seeing strings instead of point particles, it would require to go to very short distances (or very high energy scales) inaccessible at present colliders. Thus, with the currently available technology to test string theory in this context it would require building a collider of the size of our galaxy! There could certainly be surprising advances in our technology, but, notwithstanding the fact that anything is certainly possible, it is hard to be enthusiastic about such a prospect within one's lifespan.

{}The third important difficulty in string theory is that all known fully consistent string theories are supersymmetric. Supersymmetry is a symmetry between particles with half-odd-integer and integer spins. Supersymmetric theories have the property that for each particle with half-odd-integer spin there is an otherwise identical particle with integer spin. The world around us is not supersymmetric. More precisely, we do not observe supersymmetry in nature. This does not necessarily mean that supersymmetry is irrelevant in the real world -- supersymmetry could be broken just like the electroweak interaction is broken into the electromagnetic and weak interactions. In fact, supersymmetry breaking can be achieved within the context of point-like particles. What is more difficult is to construct consistent string theories with broken supersymmetry. In fact, in the context of making a connection with the real world, one invariably seems to consider supersymmetry breaking at the level of a 4-dimensional theory of point-like particles resulting from string theory rather then at the more fundamental string theory level. This, once again, diminishes the predictive power of string theory.

{}As we have already mentioned, there are five known consistent string theories in 10 dimensions as well as many ways of obtaining 4-dimensional theories from them. Could it be that the aforementioned difficulties are due to the fact that we have too many string theories to begin with? In other words, could it be that we must first unify various string theories into a more fundamental framework? This idea was successfully pursued (at least to some extent) in the 1990s. It indeed appears that the known five string theories in 10 dimensions are not all so different but are related to each other via a web of dualities. That is, some complicated phenomena in one string theory often have a much simpler description in another string theory. Moreover, various string theories, which are 10-dimensional, appear to be related to an 11-dimensional theory called M-theory. In M-theory strings are not the fundamental building blocks. Rather membranes appear to be more fundamental, albeit M-theory is still not very well understood, so what the true fundamental building blocks of M-theory are is still somewhat unclear.

{}The fact that even strings might not be most fundamental suggests that perhaps one should consider all possible extended objects to begin with. Such objects are referred to as branes -- this name comes from the name membrane. More informatively, one refers to point-like particles as zero-branes since they have no extended space-like directions; strings are one-branes as they have one extended space-like direction; membranes are two-branes, {\em etc}. It is not easy to construct a consistent fundamental theory of various branes, in particular, their interactions are difficult to describe. Zero-branes, that is, point particles are relatively easy to deal with. This is because point particles interact at a point in space-time, so their interactions are local. Locality of interactions is a key simplifying feature. Now, one-branes, that is, strings are harder -- their interactions are non-local. However, there is something very special about strings, which is related to the fact that you can draw strings on a plane. This can be understood via an analogy -- solving geometry problems on a plane is much easier than in 3 or higher dimensions. At any rate, we know (more or less well) how to deal with interacting strings, while interactions of higher dimensional extended objects are much less understood.

{}Despite the aforementioned issues, branes seem to be an important ingredient of string theory itself. Thus, in various 10-dimensional string theories there are present branes of different dimensionalities. In particular, branes play a crucial role in relating string theories via the web of dualities. This important role of branes can be understood as follows.

{}Thus, there are two types of strings -- closed strings and open strings. A closed string is a loop -- it has no ends; an open string, on the other hand, has two ends. Closed strings propagate in all 10 dimensions of string theory. However, open strings need not move in all 10 dimensions. One can have open strings that end on a lower dimensional brane (called a D-brane), so these open strings move only in those dimensions filled by the brane. Thus, consider a 3-brane with 3 extended space-like dimensions. If both ends of an open string are attached to this brane, then this string as well as all of its vibration modes (that is, the corresponding point particles) move in 4 dimensions (3 space-like directions plus time) -- this is unlike closed strings, which move in all 10 dimensions. Branes, therefore, provide a concrete mechanism for trapping (or localizing) point particles coming from string theory on subspaces which are lower than 10-dimensional. This is certainly good news -- this opens up a possibility of having the 4-dimensional world, including the point particles corresponding to the matter and interaction quanta in the Standard Model, coming out of the vibration modes of an open string ending on a 3-brane. Since these point particles are trapped on the 3-brane, we do not see the extra 6 dimensions, even if they are infinite, by doing experiments with, say, photons -- they are trapped on the brane. Can we, however, detect the extra dimensions? After all, it is precisely these extra dimensions that we would like to see to connect string theory to the real world. To understand this, let us dig a bit deeper.

{}An important fact that we need here is that open strings produce massless modes with spin not larger than 1. This is exactly the type of point particles we need to potentially accommodate the Standard Model (or its unified extensions). However, what about gravity? Gravitons carry spin 2. They come from closed strings. One way to understand this is to note that two open strings can be glued into a closed string. This helps visualize why closed strings produce spin 2 massless particles. So, the picture that emerges from this is that the Standard Model particles are trapped on a 3-brane (and therefore propagate only in 4 dimensions in accord with the experimental observations). However, since closed strings propagate in all 10 dimensions, gravitons also move in these 10 dimensions. This scenario is referred to as Brane World.\footnote{\, This term was coined in the paper:\\ Z. Kakushadze and S.-H.H. Tye, ``Brane World", {\em Nucl. Phys.} {\bf B548} (1999) 180-204; arXiv:hep-th/9809147.} In this scenario gravity and other interactions are unified via string theory, but they do not quite come on the same footing as they move in spaces with different numbers of dimensions (the 3-brane is a subspace of the full 10-dimensional space).

{}One of the immediate challenges of Brane World is that gravity lives in 10 dimensions. We observe 4-dimensional laws of gravity, not those of a 10-dimensional world. Can we then reconcile the Brane World scenario with observation? Actually, there are (at least) two known ways for this. First, we could always assume that the extra 6 dimensions are curled up into small circles (or other more complicated spaces). Then, if these dimensions are small enough, we could not detect them in current gravitational experiments. Note that since the Standard Model particles do not propagate in these extra dimensions, it is precisely gravity that we would need to employ to detect these extra dimensions. Here one can ask: How is this any different from the more traditional string paradigm we discussed above? After all, getting rid of the extra dimensions in this way was considered somewhat of a drawback. The point is that the three fundamental interactions other than gravity have been measured at very small distances, about 15 orders of magnitude smaller than a millimeter. On the other hand, gravity due to its weakness compared with the other three interactions has been measured more or less precisely only down to about 1/10th of a millimeter! That is, the extra dimensions could be as large as about 1/10th of a millimeter and we could not have known about them. But there is an exciting prospect that they could be detected in the upcoming precision measurements of gravity beyond a millimeter. This is certainly exciting; however, there is a very pressing question with this scenario -- why would the extra dimensions be precisely around a millimeter where we could detect them readily within some number of years (or decades)? Put differently, it would be one amazingly lucky coincidence if that were the case. Moreover, all the previously mentioned difficulties with compactification still remain in this scenario.

{}There is, however, another interesting way of solving the above issue with gravity in the Brane World context. Imagine that the extra dimensions are infinite. Naively, then, gravity would appear to be 10-dimensional in a gross contradiction with the experiment. However, imagine that we could have some gravitons (almost) localized on the very same brane where the Standard Model particles live. Then we have the following scenario. At short distance scales gravity would appear 4-dimensional due to these almost localized gravitons -- this is because we live on the brane, so these almost localized gravitons would dominate over those that freely propagate in 10 dimensions. On the other hand, as we go to larger distance scales, the latter become more important, and gravity becomes higher dimensional. With the appropriate choice of parameters in the theory, it appears to be possible to have this cross-over scale from 4-dimensional gravity to higher dimensional gravity as large as or larger than the observable size of the universe. So discovering extra dimensions in this scenario would involve astrophysical experiments, not collider ones. And there is a possibility that the upcoming satellite (and other) experiments of this type could discover extra dimensions via measurements at cosmic scales!\footnote{\, Recently, a third approach to extra dimensions in Brane World was proposed.  In this new scenario, gravity in 10 dimensions is massive, while gravitons (almost) localized on the brane are (almost) massless.}

{}One question that needs to be clarified in this scenario is how to get these almost localized gravitons, in particular, in string theory. It appears that one can indeed have such gravitons in the string theory context, although one is still far from building a fully realistic model with such features. Nonetheless, the Brane World scenario has opened up new avenues for perhaps very surprising ways of unifying all fundamental interactions. In fact, it also provides an exciting new possibility for addressing the so-called Cosmological Constant Problem.

{}The essence of this problem, which is thought of as the Holy Grail of theoretical physics, is the following. In quantum mechanics even the vacuum is not entirely ``empty" -- there are certain oscillations in the vacuum that contribute to the energy density of the universe. The natural value of this energy density computed on theoretical grounds is in a gross contradiction with the experiment -- it is roughly 60 orders of magnitude larger than the observed experimental bound. Why is the observed vacuum energy density, which is referred to as the cosmological constant, so small compared with its theoretical prediction? There is no satisfactory answer to this question within the 4-dimensional theory of point particles.

{}However, Brane World with infinite extra dimensions seems to give a new way of addressing this problem. The key point here is related to the fact that non-vanishing vacuum energy density curves space via its interaction with gravity, while the observed 4-dimensional world around us appears to be essentially flat. In a 4-dimensional theory large vacuum energy density we obtain from theoretical computations would curve the space way beyond experimental bounds. In the Brane World scenario, however, the space-time is not 4-dimensional but 10-dimensional. This then opens up the following new possibility -- we could have large 4-dimensional vacuum energy density on the 3-brane without curving the 4-dimensional space-time. Rather, this space could still be flat (or almost flat), while it is the extra 6 dimensions that are highly curved due to the 10-dimensional nature of gravity. Thus, the effect of large 4-dimensional vacuum energy density is essentially diluted in extra dimensions. One rough visual analogy that comes to mind is the following. Imagine a large pool of clean water. If you pour a drop of ink into this pool, it will still appear to be clear as the effect of this small amount of ink is going to be diluted -- it is just like a drop in the ocean.

{}There are still many open questions in Brane World. However, Brane World has opened up new avenues for addressing many old questions. There have been a lot of interesting developments in this exciting direction, and one can expect that there are many more to come. Hopefully, time will tell if the real world is a Brane World.

\begin{figure}[ht]
\centerline{\epsfxsize 6.truein \epsfysize 4.truein\epsfbox{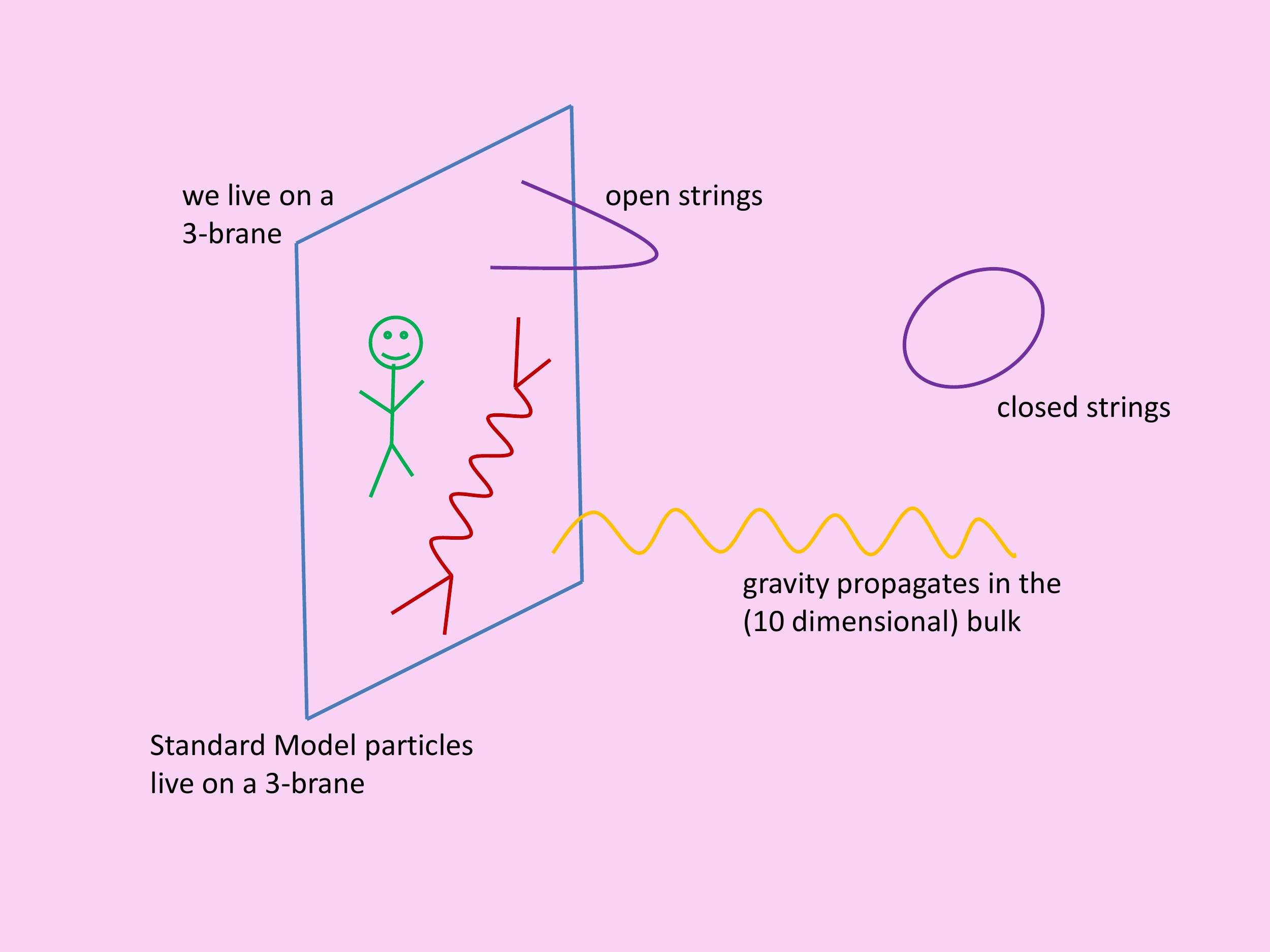}}
\noindent{\small {Figure 1. Brane World schematically.}}
\end{figure}

\end{document}